%% file: main.tex
\documentclass[conference,compsoc]{IEEEtran}

\ifCLASSOPTIONcompsoc
  \usepackage[nocompress]{cite}
\else
  \usepackage{cite}
\fi

\ifCLASSINFOpdf
\else
\fi
\usepackage{cite}
\usepackage{amsmath,amssymb,amsfonts}
\usepackage{algorithmic}
\usepackage{array,multirow,graphicx}
\usepackage{textcomp}
\usepackage{listings}
\usepackage{xcolor}
\usepackage{caption}
\usepackage{listofitems}
\usepackage{adjustbox}
\usepackage{enumitem}
\usepackage{tikz}
\usepackage{pgfplots}
\usepackage{makecell}
\usepackage{pifont}
\usepackage{booktabs}
\usepackage{subcaption}
\usepackage{float}
\usepackage{color, colortbl}
\usepackage{nicematrix}
\usepackage{comment}
\usepackage{bm}

\usepackage{url}

\usepackage{breakurl}
\usepackage[breaklinks]{hyperref}
\definecolor{grey1}{gray}{0.91}
\def\code#1{\texttt{#1}}

\definecolor{dgray}{gray}{0.25}
\definecolor{strings}{RGB}{0,0,128}
\definecolor{backcolour}{rgb}{0.95,0.95,0.92}
\definecolor{keywords}{RGB}{127,0,85}
\definecolor{darkred}{RGB}{139,0,0}
\definecolor{darkyellow}{RGB}{204,204,0}
\definecolor{darkblue}{rgb}{0.0, 0.0, 0.55}
\definecolor{vividviolet}{rgb}{0.62, 0.0, 1.0}
\definecolor{fuchsia}{rgb}{1.0, 0.0, 1.0}
\definecolor{shockingpink}{rgb}{0.99, 0.06, 0.75}
\definecolor{darkBlue}{RGB}{0, 0, 205}
\definecolor{darkgreen}{RGB}{0, 205, 0}
\definecolor{gray85}{gray}{0.85}
\newcommand{\diffadd}[1]{{\color{darkgreen}#1}}

\begin{document}

%\title{``People Actually Care About My Projects'': A Deep Dive Into How Open-Source Project Maintainers Review and Resolve Bug Bounty Reports}
\title{A Deep Dive Into How Open-Source Project Maintainers Review and Resolve Bug Bounty Reports}

\author{\IEEEauthorblockN%{Anonymous author(s)}
{Jessy Ayala, Steven Ngo, Joshua Garcia}
\IEEEauthorblockA{University of California, Irvine}
%Atlanta, Georgia 30332--0250\\
%Email: http://www.michaelshell.org/contact.html}
%\and
%\IEEEauthorblockN{James Kirk}
%\IEEEauthorblockA{School of Electrical and\\Computer Engineering\\
%Georgia Institute of Technology\\
%Atlanta, Georgia 30332--0250\\
%Email: http://www.michaelshell.org/contact.html}
}

\maketitle

\begin{abstract}
Researchers have investigated the bug bounty ecosystem from the lens of platforms, programs, and bug hunters. Understanding the perspectives of bug bounty report reviewers, especially those who historically lack a security background and little to no funding for bug hunters, is currently understudied. In this paper, we primarily investigate the perspective of open-source software (OSS) maintainers who have used \texttt{huntr}, a bug bounty platform that pays bounties to bug hunters who find security bugs in GitHub projects and have had valid vulnerabilities patched as a result. We address this area by conducting three studies: identifying characteristics through a listing survey ($n_1=51$), their ranked importance with Likert-scale survey data ($n_2=90$), and conducting semi-structured interviews to dive deeper into real-world experiences ($n_3=17$). As a result, we categorize 40 identified characteristics into benefits, challenges, helpful features, and wanted features. We find that private disclosure and project visibility are the most important benefits, while hunters focused on money or CVEs and pressure to review are the most challenging to overcome. Surprisingly, lack of communication with bug hunters is the least challenging, and CVE creation support is the second-least helpful feature for OSS maintainers when reviewing bug bounty reports. We present recommendations to make the bug bounty review process more accommodating to open-source maintainers and identify areas for future work.
\end{abstract}

\IEEEpeerreviewmaketitle
\input{content/introduction}
\input{content/related_work}
\input{content/methodology}

\input{content/results}
\input{content/discussion}
\input{content/conclusion}

%\begin{thebibliography}{1}

%\bibitem{IEEEhowto:kopka}H.~Kopka and P.~W. Daly, \emph{A Guide to \LaTeX}, 3rd~ed.\hskip 1em plus 0.5em minus 0.4em\relax Harlow, England: Addison-Wesley, 1999\end{thebibliography}

\bibliographystyle{plain}
\bibliography{references}

\newpage % The Meta-Review should at least start on a new column

% Use \appendices and not \appendix due to IEEEtran.cls quirks
\appendices % if not used earlier

\section{Meta-Review}

The following meta-review was prepared by the program committee for the 2025
IEEE Symposium on Security and Privacy (S\&P) as part of the review process as
detailed in the call for papers.

\subsection{Summary}
This paper investigates the experiences of open-source maintainers on bug bounty platforms through three studies: two surveys and interviews. Based on these studies, the authors identify and categorize 40 characteristics, finding that private disclosure and project visibility are the most important benefits, and provide recommendations for more accommodating bug bounty review processes for open-source maintainers.

\subsection{Scientific Contributions}
\begin{itemize}
\item Provides a Valuable Step Forward in an Established Field
\item Establishes a New Research Direction
\end{itemize}

\subsection{Reasons for Acceptance}
\begin{enumerate}
\item This paper provides a valuable step forward in an established field. Namely the perspectives and ranked characteristics of open-source maintainers who conduct bug bounty reviews. A step forward in the research field of software experts' perspectives concerning security bug bounties.
\item This paper establishes a new research direction. It builds on a series of recent studies investigating motivations for participating in the bug bounty market but considers a new population: open-source maintainers.
\end{enumerate}

\end{document}

%% file: content/introduction.tex
\section{Introduction}
Open-source software (OSS) has been an integral part of the software supply chain and has seen an increase in visibility and attention over the past few years \cite{hendrick2023openssf, hughes2022zdnet-oss, cisa2023sbom, glen2022smallOSS}. A 2023 Synopsys report on open-source security looked at over 1,700 code bases across a variety of industries, ranging from transportation to healthcare to enterprise software, and found that 96\% of scanned codebases contained OSS, with 76\% of the code being OSS \cite{synopsys2023report}. Lawson \& Hendrick surveyed 916 organizations across the globe and revealed that 90\% of them adopted OSS moderately \cite{world2023spotlight}. 

%\steven{candidate for removal for space purposes}
All sizes of OSS projects have a role in the OSS ecosystem, whether they are highly visible products, direct dependencies, or transitive dependencies \cite{dancuk2024dependencies}. In 2016, a developer unpublished an NPM package named \code{left-pad} that was only 17 lines of code \cite{left-pad}, which then proceeded to break thousands of other OSS projects that relied on \code{left-pad}, including Node and Babel \cite{glen2022smallOSS}.

As a result of the widespread usage of OSS across every aspect of society, there is an even greater need to ensure their security. The 2023 Synopsys report further showed that 84\% of the codebases contained at least one vulnerability and 48\% of the codebases contained a high-risk vulnerability \cite{synopsys2023report}. Codebases affiliated with the retail and eCommerce industry were shown to have had a 557\% increase in high-risk vulnerabilities over the past five years, with many other sectors seeing increases in the 200\% and 300\% range. A 2021 report from Mend on open-source security vulnerabilities revealed there were 9,658 vulnerabilities in OSS in 2020, which was a 50\% increase from 2019 \cite{murray2021mend}. 

%\steven{personal note: as Josh previously mentioned, good candidate for removal for space purposes}
The United States government recognized and drew attention to the software supply chain security issue by establishing their Open-Source 
%\josh{insert the word ``Software" here since the 3 in OS3I includes ``Software" when the acronym is spelled out} \steven{added} 
Software Security Initiative (OS3I) in 2021 \cite{oncd2023info_request}. They demonstrated continued interest with their Request for Information on Open-Source Software Security in August 2023, which received more than 100 responses from representatives across non-profits, industry, and academia involved in the OSS ecosystem, clearly showing a need for solutions to address the security of OSS \cite{oncd2024report}.

One possible solution to the software supply chain security issue is to leverage the expertise of countless security professionals all over the world through bug bounty programs, where anyone can perform their security testing on any software that is deemed in-scope for incentives such as cash awards, reputation, and physical merchandise \cite{akgul2023BBHperspectives, huang2016multiple, maillart2017eyes}. Some major companies such as Google and Microsoft maintain their bug bounty programs for their products \cite{google-bughunters, microsoftBBP}, but a majority of bug bounty programs are hosted through bug bounty platforms such as HackerOne and Bugcrowd \cite{hackerone, bugcrowd}. In particular, the \code{huntr} bug bounty platform \cite{huntr} has previously supported any OSS projects for bug bounty hunters to submit bug reports that address security vulnerabilities in the OSS projects' codebase \cite{sueiras2020startup-huntr}. As of November 30, 2023, they exclusively support artificial intelligence and machine learning OSS projects \cite{gendron2023protectAI,huntr2023migration}.

Prior work has investigated the benefits  \cite{maulani2023case-study, atefi2023chromium, beretas2023analysis, subramanian2020marketplaces, kuehn2014insttitutional, bhushan2022dynamics, zhao2015web-vulns, magazinius2021mapping} and challenges  \cite{walshe2022cvd, sridhar2021hackerone, shafigh2021invalid, zhao2016crowdsourced} of maintaining a bug bounty program, but there does not exist any work that studies bug bounty programs and platforms being applied to OSS. While several studies focus on the perspectives of bug bounty hunters \cite{akgul2023BBHperspectives, alexopoulos2021vuln-reporters, huang2016multiple, maillart2017eyes, ellis2022bounty, zhao2020control, li2022relational} and the security practices of OSS contributors \cite{hendrick2023openssf, wermke2022qual-OSSP,ayala2023workflowssps,ayala2024glimpse}, there exist knowledge gaps in the connection between bug bounty programs and OSS projects concerning the perspectives of OSS maintainers. This particular perspective is important since OSS project maintainers typically do not have a security background or adequate funding~\cite{checkpoint}, and vulnerability disclosure is a key component of further securing the OSS ecosystem. %\josh{I think it's important to have at least a sentence here explaining why we can't just ignore the OSS maintainers' perspectives.}

Privacy and confidentiality of communications, before a patch is disclosed, were highly valued by maintainers in our study, which Alomar et al. \cite{alomar2020bugs} do not discuss, as well as the pressure to review reports by maintainers. Further, while Akgul et al. \cite{akgul2023BBHperspectives} do not emphasize the importance of having proof-of-concept exploits or test cases for the vulnerability, our study highlights that point strongly as the third most challenging aspect in bug bounty report review and in terms of regression testing for CI/CD.
However, Votipka et al. \cite{votipka2018hackers} point out that bug bounty hunters and testers both say PoCs are needed to communicate the importance and severity of a vulnerability; in particular, disagreements about severity arise from bug bounty hunters when reports might not be reviewed, further annoying the hunters in the process.
Understanding how such issues arise is understudied in the context of the OSS ecosystem, which our work investigates.

In this paper, we conduct a mixed-methods study that systematically identifies and quantifies the factors that affect OSS project maintainers’ involvement in conducting bug bounty report review. 
Unlike prior work, we ask OSS project maintainers to rate the importance of characteristics, e.g., benefits and challenges, and elaborate on their rankings to allow researchers and bug bounty programs to prioritize exploring efforts for making the vulnerability disclosure and review process more accessible, i.e., for those with little or no security background. To do so, we explore the following:

\begin{center}
\noindent\fbox{%
    \parbox{0.971\columnwidth}{%
        \textbf{RQ1:}
        %What are factors that OSS maintainers consider and challenges they face when conducting bug bounty review?
        What do OSS maintainers find beneficial and challenging from conducting bug bounty review? 
    }%
}
\end{center} 

\begin{center}
\noindent\fbox{%
    \parbox{0.971\columnwidth}{%
        \textbf{RQ2:}
        How important are such benefits and challenges to OSS maintainers? Why are they important? 
        %\josh{Why what? I don't think it's very clear here what the why is referring to.}\jessy{how about now?}\josh{looks good now}
    }%
}
\end{center} 

We conduct three studies, described in Section~\ref{sec:methods}, to address our RQs: a survey ($n_1=51$) to list important benefits and challenges of bug bounty report review alongside helpful and wanted features on bug bounty platforms (RQ1), a Likert-scale survey ($n_2=90$) to rank the importance of 40 coded characteristics (RQ2) from the listing study and conduct semi-structured interviews ($n_3=17$) study to contextualize overall survey results. Section~\ref{sec:participants} describes participant demographics and their backgrounds.

We present results in Section~\ref{sec:results}. In summary, OSS project maintainers find private disclosure and project visibility to be the most beneficial aspects of bug bounty report review, while hunters' %\josh{I'm not sure what you mean by ``diverted" here. Diverted from what exactly?} 
focus on money or CVEs, as opposed to improving software security, and pressure to review bug bounty reports to be the most challenging. We also find that the most helpful features are incentivized reporting, 
%\josh{I'm not sure what you mean by ``fostered" here or how that connects with the example. From the study results, I know that ensuring the vulnerabilities are ``valid" are important to OSS maintainers, but I'm not sure how that may mean ``fostered".}\jessy{how does this wording sound?}
e.g., bug hunters receiving points for reporting valid vulnerabilities, and security metric calculation assistance. Further, we find collaboration features, e.g., the ability to request assistance from security experts, and more hunter reputation features to be the most wanted capabilities on bug bounty platforms. 

In Section~\ref{sec:discussion}, we discuss implications, i.e., based on significant challenges and benefits, for stakeholders involved in the bug bounty lifecycle and future work that encourages researchers to leverage emerging technologies, e.g., large language models, for further supporting OSS project maintainers to conduct efficient and effective bug bounty review. Finally, we conclude in Section~\ref{sec:conclusion}.

%% file: content/related_work.tex
\section{Related Work}
We look at previous work through the following lenses: (1) the benefits and challenges that come with maintaining a bug bounty program, in addition to how they have been evaluated, (2) the challenges and dynamics that come with interacting with bug bounty hunters, and (3) the current security practices of OSS project contributors. 
  
%\subsection{Benefits, Challenges, and Evaluation of Bug Bounty Programs}
%We first sought to understand the benefits of bug bounty programs, their challenges, and how they are evaluated. In particular, we want to know what aspects of successful or unsuccessful bug bounty programs stand out the most. 

\noindent\textbf{Bug Bounty Program Benefits and Challenges}.
Several studies highlight the benefits of maintaining a bug bounty program, which includes their role in creating an improved security posture within an organization \cite{maulani2023case-study, atefi2023chromium, beretas2023analysis, zhao2016crowdsourced}, responsible vulnerability disclosure \cite{subramanian2020marketplaces}, and increased number of reported vulnerabilities \cite{kuehn2014insttitutional, bhushan2022dynamics, zhao2015web-vulns}. 
%\josh{possible from?} \steven{addressed} 
Walshe \& Simpson demonstrate that, on average, maintaining a bug bounty program has a comparatively lower yearly cost than hiring two additional software engineers \cite{walshe2020empiricalBBPs}. How to successfully incentivize hackers while keeping total costs low and avoiding a diverted focus from security to financial incentives has also been studied through a game model \cite{hou2023temporal} and analyzing participation and payout rates \cite{finifter2013vuln-rewards}. Having a bug bounty program also comes with challenges, primarily dealing with invalid, low-quality reports from bug bounty hunters \cite{walshe2022cvd, sridhar2021hackerone, zhao2016crowdsourced}. However, work exists centered around mitigating such issues \cite{shafigh2021invalid, nagwani2021slr-dupes}.

%\steven{could be condensed or removed for space purposes}
Bug bounty programs have also been studied under additional contexts. Magazinius et al. conducted a systematic mapping study of research done on bug bounty programs and found a need for more qualitative studies to understand what drives product owners \cite{magazinius2021mapping}. The rules, policies, and business models of bug bounty programs have been evaluated to determine how to best cultivate an ethical hacking environment with responsible disclosure processes \cite{zhao2017policies, elazari2018policy, laszka2018engagement, ruohonen2018web}. Schulz compared statistics from penetration testing companies and web applications listed on Bugcrowd, a bug bounty program platform, to determine cost-benefit trade-offs between the two options for security evaluation \cite{schulz2014pentest}. Luna et al. emphasized the importance of considering target user populations and incentive mechanisms of bug bounty programs when making comparisons \cite{luna2019patterna}.

% \steven{Isn't really relevant to our work, just talks about bug bounty programs being applied to blockchain/IoT}
% Theoretical economic and game models have been applied to a bug bounty program context to determine potential optimal design balancing incentives, number of participating individuals, and quality of reports \cite{laszka2016banish, gersbach2023decentralized}. There exists a proposal for a gamified bug bounty program implementation built using gamification elements such as leaderboards and competition to alleviate financial resource demands \cite{ohare2020game}. Subfields of blockchain \cite{marcavage2023blockchain, bhosle2023bug-blockchain} and Internet-of-Things \cite{ding2019iot} have also had applications of bug bounty programs to enhance vulnerability management, with blockchain technology also being integrated into existing bug bounty program infrastructure to improve responsible disclosure and compensation \cite{badash2021blockchain, breidenbach2017hydra}. 

%\subsection{Challenges and Dynamics of Bug Bounty Hunters}
%We then looked at work that focused on the perspectives of bug bounty hunters, who have also been referred to as white-hat hackers and security researchers throughout the literature. 
\noindent\textbf{Challenges and Dynamics of Bug Bounty Hunters}. Akgul et al. conducted various surveys and interviews to understand the bug bounty hunters' motivations and challenges when participating in the bug bounty ecosystem \cite{akgul2023BBHperspectives}. They found that rewards and learning opportunities drive them, while communication problems demotivate them the most. Several other studies have investigated how bug bounty hunters operate, particularly what incentivizes them to participate in bug bounty programs \cite{alexopoulos2021vuln-reporters, huang2016multiple, maillart2017eyes}. Two studies have investigated how to best collaborate with bug bounty hunters to achieve effective program performance through the perspective of control theory and relations governance mechanisms \cite{zhao2020control, li2022relational}. Ellis \& Stevens provide significant socio-technical insights into bug bounty hunters, including discussions on the risks and insecurities they face as essentially gig workers \cite{ellis2022bounty}.

%\subsection{Security Practices of OSS Project Contributors}
\noindent\textbf{Security Practices of OSS Contributors}. Our understanding of OSS project contributors is rooted in work that has investigated their existing security practices and culture. 
Hendrick and Ramaswami surveyed 441 OSS contributors, 90 of whom identified as maintainers, to gather insights on how they follow and incorporate secure software development best practices in their projects \cite{hendrick2023openssf}. Unlike contributors who only make occasional contributions, maintainers are deeply familiar with their OSS projects and actively shape their future releases. Some key takeaways from their findings of the maintainers' perspectives are that 69\% of OSS contributors desire clearly defined secure software development best practices, 27\% of maintainers define their project's OSS security policy, and 30\% of maintainers implement such policy. Wermke et al. conducted 27 semi-structured interviews with OSS contributors to determine how researchers can better support them in their security and trust practices \cite{wermke2022qual-OSSP}. One important finding was the need for additional consideration for the diversity of OSS project size, their amount of resources, utilization of security and trust processes, and underlying motivations. 
%\josh{Please add a sentence here explicitly stating how our work is different from the aforementioned ones in this paragraph.} \steven{addressed} 
While our work indirectly addresses OSS contributors in the form of bug hunters, we specifically focus on maintainers. %\josh{So, you focus on both, or did you mean that you actually focus on maintainers?} \jessy{reworded}

%\josh{This paragraph does not seem so important to me and may be a target for cutting later if we have a space issue.} \steven{The content could probably be moved to Introduction to highlight importance of OSS then}
% Adapted to Introduction
% Throughout our coverage of related works, we noticed that while the perspectives of the bug bounty hunters have since been studied, the perspectives of reviewers have not yet been taken into consideration. In addition, OSS and software supply chain security have become increasingly important over the past few years \cite{hendrick2023openssf, hughes2022zdnet-oss, cisa2023sbom, glen2022smallOSS}. The United States government drew worldwide attention to the software supply chain security issue with the creation of the Open-Source Software Security Initiative (OS3I) in 2021 \cite{oncd2023info_request}. They demonstrated continued interest in their Request for Information on Open-Source Software Security in August 2023, and it received 100+ responses from representatives across nonprofits, industry, and academia involved in the OSS ecosystem \cite{oncd2024report}. 

There is a critical need to understand the perspectives of the people who develop and maintain OSS projects. Bug bounty programs are one avenue for improving the security posture of OSS projects through crowdsourced work from bug hunters, and our work is motivated by wanting to understand the role and insights of OSS project maintainers in this ecosystem. 
%\josh{Please state in this paragraph how this paper/work is novel compared to the aforementioned work/papers.} \steven{addressed} 
Our work is directly inspired by Akgul et al.'s study of bug bounty hunter perspectives \cite{akgul2023BBHperspectives}, but we instead study the other angle of the bug bounty report interaction, i.e., those who review bug bounty hunters' reports. 
%In addition, our work further differentiates itself by focusing on one bug bounty program, \code{huntr} \cite{huntr}, that has previously supported any OSS projects that chose to be listed on their platform \cite{sueiras2020startup-huntr, gendron2023protectAI}.

%% file: content/methodology.tex
\section{Methodology}\label{sec:methods}

We adopt the methodology of a recent paper that investigated bug bounty hunter perspectives \cite{akgul2023BBHperspectives}, i.e., those who submit bug bounty reports, to explore the perspectives of OSS project maintainers, i.e., those who review bug bounty reports for OSS projects.
We designed and conducted three studies to investigate our research questions. An initial listing survey study to determine characteristics (RQ1), a Likert-scale rating survey study (RQ2), and an interview study. % @Josh RQ2 is purposely referred to twice, please see the RQs at the end of Section 1 to see how they relate
The first two studies allow us to understand the benefits, challenges, helpful features, and needs of OSS project maintainers. We conduct a follow-up interview study to contextualize results.

%Our institutions’ review boards approved our study, and identifiable data was only available to authors listed on the corresponding study team tracking log. 

\subsection{Ethical considerations}

%All data was found in public spaces, e.g., GitHub profiles. As a result, we skipped recruiting OSS project maintainers if their contact information was not immediately publicly available. Our institution’s ethics review board approved all three studies. Participants signed consent forms detailing study plans and participant rights before data collection. Further, our study is GDPR-compliant.

Our institution’s ethics review board approved all three studies. Participants signed consent forms detailing study plans and rights before data collection. Further, our study is GDPR-compliant. We worked with our institution’s IRB to employ any changes that would help reduce complaints or negative feedback from potential subjects. 
For our recruitment, we identified potential subjects from the \texttt{huntr} platform and manually visited their respective project pages.
We refrained from using commits as a source of information for author emails, and instead, skipped over potential subjects if their contact information was not easily accessible, e.g., from their GitHub profile. 
This approach follows the recommendation Utz et al. \cite{utz2023privacy}: ``for future recruitment of study participants we recommend, as also suggested by the DPA, to only use contact information that has visibly been made public by the individuals themselves with the intention of allowing the general public to contact them.''

\subsection{Listing survey study}\label{sec:lss}

To identify characteristics, e.g., benefits, when performing bug bounty report review, we conducted an online survey between January 2024 and March 2024 on OSS project maintainers who have validated vulnerabilities using the \code{huntr} bug bounty platform ($n_1=51$).

\subsubsection{Participant recruitment and piloting}

We reached out to OSS project maintainers who have validated vulnerabilities via bug bounty report disclosure using the \code{huntr} bug bounty platform from July 2021 to December 2023, covering 558 GitHub projects, and had publicly available contact information in their GitHub security policy or associated with the owning GitHub account.
In particular, \code{huntr} originally supported any GitHub project. Protect AI acquired \code{huntr} in mid-2023 and offboarded projects from the platform that were not AI/ML-related; thus, our targeted subjects cover general OSS projects.

We piloted the survey with ten respondents and reviewed the quality of responses, i.e., making sure the instructions were clear to create a list for each free-response question, for the listing survey study described in Section \ref{sec:lss}. We then made wording updates and indicated to try to list at least three different responses per free-response question for categories shown in~\autoref{tab:bigasstable}.

\subsubsection{Survey details}

We asked participants to list characteristics associated with conducting bug bounty report review in four categories: benefits of bug bounty review, challenges of bug bounty review, helpful platform features, and wanted platform features. Next, we asked participants to self-report their OSS maintenance and industry experience, if they have a security background, project funding, and if reviewed bug bounty reports resulted in a CVE. We ended with demographic questions to understand our population.

\subsubsection{Data analysis}

We analyzed listing survey responses with exploratory open-coding~\cite{saldana2021coding}.
Two researchers independently coded batches of ten responses at a time; further, resolving differences and updating the codebook after each batch.
Because the survey lent itself to multiple listings per category, e.g., benefits, % shown in~\autoref{tab:listings}, 
we were able to create an initial codebook from 18/51 (35.3\%) respondents, resulting in 29/40 (72.5\%) characteristics shown in~\autoref{tab:bigasstable}. 
\begin{comment}
\begin{table}[htbp]
\caption{Listing Study Responses and Codes by Category}
\label{tab:listings}
\begin{center}
    \begin{tabular}{p{0.22\columnwidth} p{0.25\columnwidth} p{0.28\columnwidth}} 
 \hline
\textbf{Category} & \textbf{\# of coded listings} & \textbf{\# of resulting codes}\\
\hline
Benefits & 122 & 10\\
\hline
Challenges & 120 & 12\\
\hline
Helpful features & 78 & 9\\
\hline
Wanted features & 44 & 9\\
\hline
\end{tabular}
\end{center}
\end{table}
\end{comment}
In particular, two researchers analyzed the same 18 participant responses by engaging in open coding~\cite{charmaz2014constructing} and discussing the initial emerging themes, meeting four times. Both researchers then independently coded the remaining 24 responses in batches of 4 and met frequently to discuss findings and reach consensus, consequently resulting in 11 emerging codes after the initial analysis. The final 8 responses were received after the final codebook was established, containing 40 characteristics across the four categories shown in~\autoref{tab:bigasstable}.

\subsection{Likert scale survey study}

The listing study identified a thorough set of characteristics considered by OSS project maintainers, but not their relative importance. Thus, we conducted a second study between February 2024 and May 2024 asking participants to rate how important each characteristic is to them using a five-point Likert scale ($n_2=90$).

\subsubsection{Participant recruitment and piloting}

We reached out to OSS project maintainers who have validated vulnerabilities using the \code{huntr} bug bounty platform from July 2021 to January 2024, covering 612 GitHub projects, and had publicly available contact information in their GitHub security policy or the owning GitHub account. We also expanded our recruitment to include OSS projects that have used third-party delegated bounty programs, such as 
%IssueHunt~\cite{issuehunt}, which provides financial support to open-source projects and their contributors through issue-based bounties., and 
Bountysource~\cite{bountysource}, which was a crowdsourcing website for open-source bounties, and OSS projects that have used bug bounty programs generally, such as HackerOne~\cite{hackerone}, which also offers a version of their popular HackerOne Bounty program for free to eligible open-source projects; thus, resulting in  
an additional 509 OSS projects covered. We allowed participants to submit responses during and after conducting interviews in hopes of receiving more responses.
Further, roughly 20\% of Likert survey recruited participants are from the Listing survey, so a majority of the Likert survey participants are likely new participants, thereby minimizing the effect of bias for the Likert survey study.

Before survey deployment, we asked personal connections in our department who maintain an OSS project to review the definitions in our study to gauge their understanding of the 40 characteristics identified from the listing survey study described in Section \ref{sec:lss}. We made wording updates and added an example when a definition was unclear to ensure survey respondents could accurately understand the intended meaning of each characteristic.

\subsubsection{Survey details}

Survey participants were asked to rate each of the 40 characteristics on a five-point Likert scale, which are shown as rows in~\autoref{tab:bigasstable}. Again, we ended with standard demographic questions to understand our sample population.

\subsubsection{Data analysis}\label{likertanal} We adopt comparison-based techniques that consider whether one characteristic is rated higher than another because using traditional statistical techniques, e.g., averages, has been criticized and discouraged by statisticians \cite{gitta2004applying,dittrich2007paired,Wu2007AnES} due to the assumption that the ordinal options in a Likert-scale are equidistant. 
In particular, we employ Log-Linear Bradley-Terry (LLBT) modeling to synthesize worth estimates ($\pi$) that represent the relative importance of characteristics to participants on a preference scale \cite{Dittrich2009FittingLB}. 
The probability of one characteristic, e.g., a benefit, being preferred over another is given by:
\begin{center}
$p(f_j > f_k|\pi_j,\pi_k) = \frac{\pi_j}{\pi_j + \pi_k}$
\end{center}
where $j, k$ denote the indices of factors considered~\cite{Dittrich2009FittingLB}. 
%From this point forward, we rename worth, $\pi$, to \textbf{RS}, ranking score, for consistency. \josh{I think you should just use $\pi$ or \textbf{RS} throughout the paper, not both.}

\subsection{Interview study}

We asked consenting and interested listing study participants to participate in a remote semi-structured interview between February 2024 and May 2024 to learn more about why the identified factors they listed were especially important to them ($n_3=17$).

\subsubsection{Participant recruitment and piloting}

Interviewees were a subset of the initial Listing survey study described in Section \ref{sec:lss}. We invited all eligible---i.e., we required that (1) the participant is an OSS maintainer and (2) the participant has reviewed more than 2 (median) bug bounty reports because we wanted to capture various reviewing experience levels---and interested listing study participants to participate in remote semi-structured interviews. 
%We directed them to a Calendly \cite{calendly} space where they could join Zoom link during specified time ranges. This also allowed us to access their corresponding non-identifiable demographic information and background as they were prompted to select if they were interested in participating in the interview study. 
24 participants originally agreed to be interviewed. After reaching out using publicly listed contact information from GitHub projects, 19 responded; afterward, 17 participants showed up. 
We conducted pilots with three participants to test our semi-structured interview protocol. Based on the quality of the interviewees' responses, we revised our interview questions to further narrow our research focus.
In particular, we made minor updates to four interview questions so that they were clearer to reduce the amount of time spent on follow-up clarifications and removed two questions entirely as they were repetitive. 
Furthermore, we iteratively updated the interview guide based on conducted interviews and participant feedback. 
Changes were limited to adding a few follow-up questions and minor structural modifications, reaching saturation without any changes past the 11th interview.

\subsubsection{Interview details}

We conducted semi-structured interviews with 17 OSS project maintainers, averaging 53 minutes. Before meeting, we emailed participants a study information sheet for review and acquired verbal consent at the beginning of each interview. Although survey participants were not compensated, interviewees were thanked with a virtual Visa \$20 gift card. We conducted interviews while survey recruitment was available to retain more participants.

We started by asking each participant about their general roles and duties for their respective managed projects, followed by why their listed benefits, challenges, helpful features, and wanted features were especially important to them. Due to time constraints and infeasibility, we refrain from asking about all 40 characteristics shown in~\autoref{tab:bigasstable}; however, we ask participants to focus on characteristics they perceived as significant, naturally leading to talking about others. We then asked about their perceptions and familiarity with the vulnerability disclosure process and experiences regarding vulnerability management in general. Finally, we asked whether and how they would consider using emerging techniques, e.g., large language models, to help improve the quality and speed of bug bounty report review.

\subsubsection{Data analysis}

All 17 audio-recorded interviews were transcribed and were checked for quality and accuracy by the researchers. Data was analyzed using thematic analysis~\cite{braun2006using}, starting with open coding~\cite{charmaz2014constructing} using each transcript, and developing thematic codes with axial coding~\cite{charmaz2014constructing} to describe common arising themes, e.g., report processing and perceptions (Section \ref{sec:rpp}). Two researchers then engaged in memo writing and constant comparison~\cite{glaser1965constant}, and inductive analysis based on grounded theory~\cite{charmaz2014constructing}. Both researchers analyzed the same 5 transcripts by engaging in open coding~\cite{charmaz2014constructing} and discussing the initial emerging themes, meeting three times.
During the analysis and iteratively discussing emerging themes across the 5 transcripts, both researchers engaged in axial coding~\cite{charmaz2014constructing}, analyzing the remaining 12 transcripts in parallel, and met frequently to discuss findings and reach consensus. 
Researchers were open to emerging themes; if a theme was adopted after discussion, researchers returned to previous interviews and re-coded accordingly.

\subsection{Limitations}

Our methodology relies on self-reported data, which often entails substantial noise. To mitigate this, we investigate our research questions using three studies, consisting of two survey studies and one interview study; meanwhile, we maximize face validity by piloting each stage of the study, and revising procedures with feedback. Further, not all of our sample participants likely have extensive experience with bug bounty report review; however, we expect some participants with less bug bounty report review participation will have similar experiences or various perspectives.

%\steven{Should we cite another paper as a comparison point? Or any other source that allows us to claim these numbers are sufficient.}
While we had sufficient participants to conduct our studies, i.e., 17 interviewees and 141 survey respondents, we cannot claim our results can be necessarily generalized to the OSS project maintainer demographic. To mitigate this, we conduct three studies to ensure a holistic perspective, and our sample population varies in range regarding education level and years of involvement in managing OSS projects, as shown in~\autoref{tab:participants}. Further, although \texttt{huntr} is our primary source of recruitment for the listing and interview studies, 558 GitHub projects are sampled and range in popularity, some with as many as 127 thousand stars. We expand this range to 509 additional projects for the Likert-scale study to further ensure a holistic range of OSS project maintainers. %~\cite{vimproject}.

%\diffrem{Our gift card incentive might have been most attractive to interview participants who make less money in OSS maintenance, perhaps due to lower skill or experience; however, }
OSS project maintainers typically are not funded nor have members that are primarily security-oriented, so our participants are likely to participate in the interest of further securing the OSS ecosystem rather than monetary motivation, i.e., survey participants are not compensated.
We addressed this by recruiting all valid listing and interview studies' participants directly through email, and Likert-scale study participants from GitHub projects of publicly disclosed reports from various bug bounty programs.

%% file: content/results.tex
\section{Participants}\label{sec:participants}

~\autoref{tab:participants} summarizes participants’ self-reported demographics and experiences. We had 51 participants in the listing study, 90 in the Likert-scale rating study, and 17 interviewees. Participants were mainly from Europe and North America and were overwhelmingly male, which is consistent with OSS developer qualitative studies, where the Americas and Europe comprise most participants \cite{huang2021fingerprints,liang2022understanding}.

\begin{table}[th!]
    \begin{centering}
    \begin{adjustbox}{width=0.48\textwidth}
	\begin{tabular}{llrrr}
            \toprule
            \midrule
            \centering
		& & \textbf{L} & \textbf{I} & \textbf{LS} \\ 
		\midrule
            \midrule
		\textbf{Gender}     & Man & 47 & 15  & 80 \\
                                & Woman & 2 & 1  & 6 \\
                                & Non-binary & 2 & 1 &  4 \\
            \midrule
		\textbf{Age}        & 18-29 & 9 & 2  & 22 \\
                                & 30-39 & 24 & 6  & 32 \\
                                & 40-49 & 10 & 6  & 23 \\
                                & 50+ & 8 & 3  & 13 \\
		\midrule
		\textbf{Residence}  & Australia & 2 & 0 &  7 \\
                                & Africa & 1 & 0 &  1 \\
                                & Europe & 31 & 11 &  41 \\
                                %& Middle East & 0 & 0 &  1 \\
                                & North America & 13 & 5 &  31 \\
                                & South America & 1 & 0 &  2 \\
                                & South Asia & 2 & 0  & 2 \\
                                & Southeast Asia & 1 & 1  & 4 \\
                                & Other & 0 & 0  & 2 \\
		\midrule
		\textbf{Education}  & $\le$ Completed high school & 3 & 0  & 4 \\
                                & Trade/technical/vocational & 3 & 1  & 2 \\
                                & College, no degree & 8 & 3  & 8 \\
                                % & Associate’s degree & 0 & 0  & 0 \\
                                & Professional degree & 3 & 0  & 8 \\
                                & Bachelor's degree & 16 & 7  & 31 \\
                                & Graduate degree & 18 & 6  & 37 \\
            \midrule
		\textbf{Years working} & $<$ 1 year & 2 & 0  & 4 \\
            \textbf{in industry} & 1-5 years & 4 & 0  & 12 \\
                                & 5-10 years & 9 & 3  & 15 \\
                                & 10+ years & 36 & 14  & 59 \\
                                %& None & 2 & 0  & 0 \\
            \midrule
		\textbf{Years in OSS} & $<$ 1 year & 0 & 0  & 1 \\
            \textbf{maintainence} & 1-5 years & 10 & 3  & 26 \\
                                & 5-10 years & 18 & 7  & 27 \\
                                & 10+ years & 23 & 7  & 36 \\
            \midrule
		\textbf{Has a security} & Yes & 13 & 3  & 28 \\
            \textbf{background} & No & 38 & 14  & 62 \\
            \midrule
		\textbf{OSS project has} & Yes & 40 & 14  & 68 \\
            \textbf{at least one CVE} & No & 11 & 3  & 22 \\
            \midrule
		\textbf{OSS project has} & Yes & 22 & 7  & 30 \\
            \textbf{sponsorship or} & No & 29 & 10  & 60 \\
            \textbf{public funding} & & & & \\
            \midrule
            \midrule
            \textbf{\# of participants} &  & \textbf{51} & \textbf{17} & \textbf{90} \\
            \midrule
            \bottomrule
        \end{tabular}
    \end{adjustbox}
    \caption{The number of participants across multiple studies along with their respective demographics, various backgrounds, and project details.  Acronyms listed represent, respectively: listing (\textbf{L}) survey study, interview (\textbf{I}) study, and Likert-scale (\textbf{LS}) survey study.}
    \label{tab:participants}
    \end{centering}
\end{table}

In the listing survey and interview studies, our participants are OSS project maintainers who have received and patched vulnerabilities from the \code{huntr} bug bounty platform. We are especially curious about this subset of OSS project maintainers because it is more likely they do not have project-hosted bug bounty programs and, instead, have utilized a third-party bug bounty platform, i.e., \code{huntr}, to review reports and have no involvement in distributing bounties to bug hunters.

For the Likert-scale survey study, we contacted the same participants initially surveyed and expanded to projects on GitHub that have used any bug bounty program in general. By doing so, we have a more holistic perspective of OSS project maintainers who have reviewed bug bounty reports, regardless of their ability to help fund bug bounties.

\section{Results}\label{sec:results}

 We identified 40 characteristics that OSS project maintainers listed after having conducted bug bounty review, falling under four categories: benefits, challenges, helpful features, and wanted features. A complete list can be found in~\autoref{tab:bigasstable}, where we have also listed relevant prior work for each characteristic, if applicable. 
 
 Most prior work has focused on characterizing the bug bounty ecosystem, bug bounty economics, and the perspective of bug bounty hunters; however, \textbf{we are the first to investigate the perspective of bug bounty report reviewers who are OSS project maintainers and have patched vulnerabilities as a result of triaged bug bounty reports}. %\josh{I guess by ``aspect" you mean one of the 40 characteristics. If so, please avoid using different terms and stick with one term, probably ``characteristic" as it appears to be the more widely used term referring to that idea in this paper.} 
 Further, \textbf{we provide the first ranking of characteristics' significance, e.g., challenges, from OSS project maintainer perspectives who have conducted bug bounty report review}. This particular perspective is important as OSS project maintainers typically do not have a security background, and vulnerability disclosure is a key component of transparency in the OSS ecosystem. %\josh{Maybe, say something as to why this particular perspective is important.} 
 %\jessy{feedback addressed}

We present how many interview participants discussed or argued a particular topic in detail (\textit{\textbf{I}}), how many listing study participants mentioned a characteristic (\textit{\textbf{L}}), and the importance of characteristics from the Likert-scale worth estimates ($\bm{\pi}$) described in Section \ref{likertanal}. The relative importance of characteristics in a category, e.g., benefits, to participants is based on their Likert-scale responses, where a greater $\pi$ value indicates the probability of one characteristic being preferred over another, i.e., $\pi \in (0,1)$ and the sum of worth estimates for characteristics in a given category is 1. 
%\josh{I think more needs to be discussed with respect to Likert-scale worth estimates. What is the range of values that it takes on? What does it mean to have a lower value or a higher value for it? Otherwise, the paper does not seem so self-contained.}
%\jessy{feedback addressed}

\input{content/original_table}

\subsection{Vulnerability disclosure}

The nature of OSS reflects open vulnerability disclosure after a patch is applied. Although all interviewees agree with this, some have had negative experiences with bug hunters, resulting in early disclosure, and some have neutral or negative feelings towards the idea of disclosure in general. OSS project maintainers also have mixed feelings about the CVE process and bug bounties in general.
For instance, in Alomar et al. \cite{alomar2020bugs}, a participant mentioned that developers might feel embarrassed or blamed if someone reports a vulnerability in their code, and argued how such attitudes can discourage developers from addressing reports properly.

\subsubsection{Private disclosure}

Though not the most mentioned benefit in the listing study, \textit{private disclosure} is the highest-ranked benefit ($L=4$, $\pi=0.155$). %\josh{Without describing $\pi$ more, I don't know what the value it takes on means. It's risky to assume all the reviewers will understand what it means.}. 
This is consistent with \textit{secure disclosure} being the most listed helpful feature, even though it is ranked fifth ($L=15$, $\pi=0.118$). 

% 1st I: Benefits OSS Users and Trust
During interviews, OSS project maintainers described the importance of ensuring that vulnerabilities were reported in private and not to be disclosed without permission or until after a fix is published. Many interviewees ($I=8$) asserted that a transparent handling of security issues builds trust with users, i.e., by disclosing the associated bug bounty report and interactions with hunters after a vulnerability is patched. 
%\josh{This sentence placed here makes it sound like you are trying to relate private disclosure as a kind of transparent handling, but how is a private disclosure transparent to users? It seems opaque to users rather than transparent to them.}.
%\jessy{feedback addressed}
Further, others mentioned how they feel positive when a bug hunter finds a security bug and reports it responsibly ($I=7$) as a part of the vulnerability lifecycle. 

% 1st I: Helpful features - Secure Disclosure
Interviewees felt that private disclosure allows the necessary time to create a patch before notifying users to upgrade ($I=3$). One interviewee mentioned how they do not mind if a bug hunter opens a GitHub issue \cite{gh_issues} to report a vulnerability. In contrast, all other interviewees ($I=16$) encourage bug hunters to reach out via email or other processes specified in security policies that allow private disclosure. In another instance, an interviewee mentioned how their project handles ``very vital data'' and ``being compromised would potentially put my job in a tricky position'' (P7), further supporting how even OSS project maintainers emphasize private disclosure before publishing a vulnerability.

\subsubsection{Experiences}\label{sec:exp}

As mentioned prior, \textit{private disclosure} is the highest-ranked benefit ($L=4$, $\pi=0.155$). Although \textit{scheduled disclosure} is the least mentioned wanted feature in the listing study and its significance is ranked last ($L=2$, $\pi=0.056$), the second most popular listed challenge is \textit{pressure to review} and is ranked second ($L=12$, $\pi=0.113$), challenging the previous observation.

Most interviewees do not mention negative experiences with the vulnerability disclosure process ($I=14$). Some believe that the bug bounty process alone should sway hunters from disclosing vulnerabilities early ($I=4$); however, few interviewees experienced early vulnerability disclosure ($I=3$), i.e., the vulnerability was disclosed before approval by the OSS project maintainer. One interviewee describes a dilemma they faced, ``we recently had a case where the component they released announced that vulnerability with the steps to follow same day as the patch and it was a day before Christmas holidays'' (P9), two others had similar experiences happen 
%\josh{I'm not sure what ``this" refers to here. I'm confident it's probably not a near-Christmas holiday patch for a component release, but I'm not sure what the general form of ``this" is like here.} 
%\jessy{feedback addressed}
on multiple occasions ($I=2$). These disclosures associate negativity with bug bounty programs and may sway maintainers from considering their usage.

Some interviewees describe bug hunters as being ``pushy'' ($I=3$), and others feel ``threatened'' ($I=2$) in some cases. 
In particular, one interviewee described, ``If I don’t reply right now, could the reporter be tempted to disclose or exploit the vulnerability they reported?'' \diffadd{(P2).} Additionally, the sense of urgency and pressure raises concerns about the impact on collaboration between bug hunters and project maintainers, who described feeling overwhelmed and having had negative experiences ($I=5$). 
Furthermore, interviewees argue how they need revised bug hunter reputation systems and guidelines to foster a more cooperative and safe environment during the report review ($I=2$). 

\subsubsection{Perspectives}

\textit{Benefits OSS users and trust}---i.e., transparent handling of security issues builds trust with users, in addition to users being able to utilize more secure software---is the third-most listed benefit and is sixth in the ranking of importance ($L=13$, $\pi=0.077$). %Many interviewees ($I= TBD$) mention putting users in the open-source ecosystem. 
%Many participants in Alomar et al. \cite{alomar2020bugs} also expressed similar positive experiences.

Most interviewees view vulnerability disclosure positively ($I=14$), which is consistent with Alomar et al. \cite{alomar2020bugs} participants' perspectives. Some describe disclosing a vulnerability after patching as ``one of the strengths of open source'' ($I=4$), as it allows dependencies and affected projects to be informed of updates as soon as possible. Further, several interviewees felt that if vulnerabilities were mishandled or it was not clear that a necessary patch was deployed, there would be a negative impact and be reflected upon the project ($I=7$). As one interviewee described, ``it is important to users to show that we are listening'' (P7).

On the other hand, some interviewees argued that they had no choice and might as well keep it transparent to show people the vulnerability was fixed, but they believed it looked bad for the respective project ($I=3$). These same interviewees also reference prior experience using vulnerable dependencies but are unaffected by the specific vulnerability stated. One interviewee stated, "clearly [this] is stupid because this doesn't affect the software" (P2), but felt obligated to apply a patch, disclose the vulnerability, and notify users to upgrade.

Interviewees reference CVE creation support as an essential feature ($I=7$) as it streamlined and assisted the creation process during bug bounty report review. Some perceive the CVE process as complicated ($I=6$), where one even described it as a ``broken system'' (P8). Another interviewee affirms that they ``don't give a shit about the CVE, I care about the problem being solved'' and feels that CVEs are a ``very weird incentive'' (P3). 

\subsection{Securing the open-source ecosystem}

All three of our studies reflect that OSS project maintainers aim to secure the OSS ecosystem ``for everyone'' through conducting bug bounty review. To do so, OSS project maintainers feel it is vital to improve their project's security posture and dedicate the necessary time and resources to resolve vulnerabilities. 

\subsubsection{Security posture}

\textit{Improved security posture} was the most popular item in our listing study and the fourth-ranked benefit from the Likert-scale study ($L=33$, $\pi=0.124$). Further, \textit{project visibility} was the third-most listed benefit and ranked second in significance ($L=13$, $\pi=0.142$) respective to other characteristics discovered, reflecting an appreciation towards exposure and openness to criticism.
Alomar et al. \cite{alomar2020bugs} also find that the usage of bug bounty programs indicates higher security posture and how wide project visibility can help discover vulnerabilities that were previously not noticed by developers.

Interviewees expressed that having ``many eyes'' on the project and involving the community in searching for vulnerabilities ($I=9$) allows vulnerability detection that is otherwise impossible for OSS project maintainers alone. In particular, one interviewee stated, ``When we implement something and maintain it for a number of years, we try to think of all possible edge cases and put ourselves in the shoes of an attacker trying to secure stuff; but it's inevitable that we miss some things'' (P9). Others also mentioned how project exposure allowed older security bugs, some as old as ten years, to be discovered and patched ($I=3$). Interviewees also recognize the importance of fixing vulnerabilities causing a ``ripple effect'' in the software supply chain ($I=3$), e.g., alerting affected projects to upgrade. 

On the other hand, others had received and reviewed reports that were deemed low-quality because of their minor contributions ($I=6$). For instance, one interviewee described how after applying a fix to a low-quality report they received, the ``real world quality of the software has essentially not really been improved'' (P11), leaving them feeling like their time was wasted and an established negative association with receiving bug bounty report notifications. Results regarding OSS project maintainers' perspectives on low-quality and spam reports are presented in Section \ref{sec:rq}.

\subsubsection{Time commitments and priorities}\label{sec:time}

As expected, bug bounty review is reported as a \textit{time-consuming} challenge but is not ranked as relatively challenging ($L=10$, $\pi=0.066$). In this context, the benefit of having an \textit{improved security posture} ($L=33$, $\pi=0.124$) reflects a need to address vulnerabilities regardless of team size and available resources. 
Further benefits such as \textit{detailed reports} ($L=8$, $\pi=0.138$) and \textit{outsourced work} ($L=4$, $\pi=0.064$) help speed up the patching process.
Alomar et al. \cite{alomar2020bugs} validate this by finding that a broader scope for bounties is likely to incur more time costs and how the lack of detailed information slows down the overall vulnerability resolution process.

Interviewees expressed struggle in bug bounty review because of time constraints ($I=7$). "We only have a certain amount of time we can spend on the whole project" (P4). ``Outside hobbies'' and ``family time'' were some of the directly stated reasons why it is challenging to resolve reports in a timely manner. %\jessy{feedback addressed}
%\josh{timely? Do you mean quickly or in a timely manner?}. 
In the context of bug bounty review, some interviewees also argued that the actual security impact influences when the report will be prioritized or investigated further based on other tasks already in progress ($I=6$).

All interviewees expressed interest in finding ways to make time for bug bounty reports with legitimate vulnerabilities ($I=17$). One, in particular, stated how ``just having the reports come into us has tightened up some areas of the code where we hadn't spent much time doing things like audits'' (P8), creating time to focus on other issues. Others mention having security in mind while writing code to save time and prevent a particular category of vulnerabilities from being reported ($I=17$), paying close attention to concepts such as ``input sanitation'' when possible.

\subsubsection{Delegated bounties and support}

Although \textit{delegated bounties}, i.e., bounties for which payments come from and are handled by an entity external to the OSS project under examination, was not the most popular listed benefit and has a lower ranking ($L=9$, $\pi=0.070$) respectively, it is what makes bug bounty support possible for OSS projects without sponsor funding. 
This observation is also supported by the high number of listings for \textit{reducing project costs}, which is ranked as the fourth most helpful feature ($L=12$, $\pi=0.121$), and the ability to provide \textit{more financial incentives} to bug hunters, which is ranked as the third most wanted feature ($L=4$, $\pi=0.145$).
Further, participants from Alomar et al. \cite{alomar2020bugs} mentioned that providing extra incentives is not common and that the current bug bounty ecosystem does not incentivize external researchers to go beyond surface-level checks.

Interviewees emphasized the importance of delegated bounties to bug hunters ($I=4$) because it ``makes [them] care more about the issue''. One interviewee stated, ``When companies give back to the community, that's nice, so I think it's giving a really good incentive'' (P1). Others desire additional or more significant financial incentives to get the attention of security researchers and experts to focus on a specific target or write more detailed bug bounty reports in general ($I=2$); thus, encouraging holistic OSS security assessments.

Some also argued the value of having a bug bounty platform administrator 
%\josh{By administrative input, do you just mean OSS project maintainer? If so, I would just use that term. Admin input can mean a variety of different things, which isn't clear what the scope is here.} \jessy{updated for specificity}
to be tagged in discussions and perform additional actions ($I=2$), e.g., opening a pull request for OSS project maintainer contact information via security policy. 
Other interviewees especially enjoyed the idea of being able to easily get in contact with OSS project maintainers and discuss those potential issues ($I=5$). However, some felt like they were being attacked by hunters ($I=4$), which we present along with the idea of having a ``software ego'' in Section \ref{sec:feelings}, and forced to create an account to address unwarranted reports. In particular, one interviewee expressed frustration with the exposure of bug bounty to their project as they argued hunters ``snoop around for a bounty in a way that they weren't before, so I worry that now there's an expectation [for a payment]'' (P11), even if the issue is reported through other means, e.g., email.

\subsection{Report processing and perceptions}\label{sec:rpp}
%\josh{This title seems overly vague and unexciting. How about something like ``Report processing"?} \jessy{updated}
Multiple participants state that the primary duty of an OSS project maintainer is to perform code review (i.e., as a part of the pull request review process). Many participants noted how bug bounty reports are just security-oriented pull requests, and feel that reports must be detailed and precise but understand that low-quality, spam, inaccurate, and duplicate reports are bound to happen in both OSS and bug bounty contexts. Some bring up the idea of having a ``software ego'' when receiving bug bounty reports and even feeling personally attacked during such initial communications. Interviewees also provide advice on how new reviewers should ``not panic'' or rush to resolve reports.

\subsubsection{Report quality}\label{sec:rq}

\textit{Detailed reports} is the third least listed benefit and ranked third ($L=8$, $\pi=0.138$) respective to other benefits. Interviewees argued its importance in aiding effective and timely vulnerability resolution ($I=8$), and participants further list how \textit{incentivized reporting}, 
%\josh{This term should be defined and not just provided an example for.}
e.g., by assigning points, is helpful for hunters to increase their reputation for submitting valid reports and is ranked as the most helpful feature ($L=4$, $\pi=0.167$). However, \textit{low-quality or spam} reports is the most listed challenge and is ranked as the fifth ($L=22$, $\pi=0.094$) largest issue. 
Alomar et al. \cite{alomar2020bugs} also revealed concerns with low-quality or noisy reports from bug bounty hunters.
Many interviewees felt that this resulted from hunters having a ``diverted focus'' ($I=10$), focusing on money or CVEs ($L=12$, $\pi=0.141$), which is ranked as the most challenging aspect of bug bounty report review.

As previously mentioned, many interviewees emphasized the importance of receiving detailed reports ($I=10$). In particular, the ``most important one is the ease of reproducibility'' (P8), which we report in Section \ref{sec:reproducibility}. About half of the interviewees ($I=8$) mentioned how detailed reports help with severity assessment, reported in Section \ref{sec:severity}, and with bug fixing and testing, reported in Section \ref{sec:testing}. ``Typically go back and ask, well, how did you do that? What steps did you follow? Can you show me a screenshot or video?'' (P16). One interviewee argues that helpful reports, e.g., having enough detail for easy resolution, should receive additional financial incentives.

Many interviewees describe the challenge of receiving low-quality reports ($I=11$). As one interviewee stated, ``understanding what the reporter wants to communicate... I didn't actually understand what they were saying'' (P8). Another interviewee described such reports as ``really drunk [and] looks like garbage'' (P15). Such reports ``pollute'' the value of bug bounty reports in general and leave OSS project maintainers in doubt about their legitimacy ($I=2$).

Interviewees argue that the mass amount of low-quality and spam reports is because of a diverted focus on bounties and CVEs ($I=10$). One interviewee told us, ``Some of those low-effort reports where you feel like they're just trying to get the bug bounty money, I think that's like the biggest annoying part'' (P4). One interviewee also argued that some bug hunters only care about receiving a CVE to improve their reputation. On the flip side, another interviewee specifically stated how ``nobody is really trying to extort us for money, which is a very common thing that occurs'' (P2),  
%\josh{Eh, nobody is extorting them for money, and that's unfortunate?}
%\josh{nobody trying to extort them for money sounds like a good thing though} \jessy{updated}
%\jessy{feedback addressed}
reflecting a positive view of bug hunters, who are further described as ``very professional'' during disclosure.

Some interviewees suspect that a subset of hunters they have encountered are at an early stage of their security career ($I=2$), e.g., college students. One interviewee describes that some bug hunters think they are ``hot shit researchers, but it was pretty apparent they were quite young... I don't want to say I was thrilled to be their teacher on this stuff [and] you know, helping them to figure it out'' (P15). 

\subsubsection{Reported severity}\label{sec:severity}

Although not the most listed challenge, \textit{overstated severity} was ranked fourth ($L=9$, $\pi=0.096$) compared to other challenges during the bug bounty report review process. However, \textit{validating impact} was the third most listed challenge and ranked fifth ($L=11$, $\pi=0.085$), respectively. 
Alomar et al. \cite{alomar2020bugs} also provide evidence that CVSS scoring and vulnerability rating mechanisms are insufficient when assessing vulnerability severity, and how misunderstood impact could result in closing reports. Further, Akgul et al.\cite{akgul2023BBHperspectives} reveal that hunters' dissatisfaction with responses, e.g., downgraded severities, are particularly challenging to deal with.
Related characteristics include \textit{feedback capabilities} for report reviewers 
%\josh{feedback to whom and from whom? please clarify here} \jessy{updated} 
as a helpful feature ($L=6$, $\pi=0.122$) and a bug hunter \textit{reputation system} ($L=4$, $\pi=0.145$) as the second most wanted feature.

Several interviewees felt that the severity reported by bug hunters was exact or close enough to the actual severity in a real-world software deployment context ($I=6$), two of which described how hunters only seemed to initially run security tooling or glance at the source code without setting it up before submitting a bug bounty report, and hunters did not ``learn the software.'' 
On the other hand, Akgul et al. \cite{akgul2023BBHperspectives} found that hunters feel unclear scopes pose additional challenges, e.g., not knowing what constitutes a valid vulnerability.
In our interpretation, understanding the source code is a core component of calculating severity because, without it, the context in which the security bug exists may overestimate its actual severity. %\josh{I'm not sure what the implications are of not setting up the software before submitting a report. I think you should elaborate.}

% 1st I: Overstated or inaccurate severity
A majority of interviewees argued how they received reports where the severity was ``overstated'' compared to the actual security impact on the software ($I=11$). Further, interviewees noticed that these reports consisted of attack vectors assuming administrative privileges ($I=7$). One interviewee described such attacks as ``shooting themselves in the foot'' and not a priority since ``admin isn't likely to sabotage their own piece of software'' (P11).

In other scenarios, some argued that the bug hunter's initial severity was almost useless and only accurate in some cases ($I=2$). Stemming from the idea that bug hunters are motivated by bounties, one interviewee explained how reports were ``almost guaranteed to be high severity and people would just check off all the maximums to see if they could get that through'' (P2) and receive a higher bounty or CVE as a result. Interviewees felt that being able to give feedback on the severity and performing modifications was a beneficial feature ($I=7$). Further, having the ability to review or assign a credibility factor to bug hunters is a wanted feature by OSS project maintainers.

\subsubsection{Duplicate reports}

\textit{Duplicate reports} was the second least listed challenge and ranked sixth ($L=6$, $\pi=0.043$). Though duplicate reports seem not to be a significant issue for OSS project maintainers, receiving them can be ``frustrating'' and ``a waste of time to deal with'' ($I=7$).
Akgul et al. \cite{akgul2023BBHperspectives} found that bug bounty hunters find duplicate reports particularly challenging, showcasing opposing opinions on the value of duplicate reports between these reviewers and hunters.
%\josh{I don't think the ending clause in the previous sentence makes sense. How about this? ``..., showcasing opposing opinions on the value of duplicate reports between these reviewers and hunters''.}}

One interviewee found that the best way to deal with duplicates is to simply ignore them. However, others noted that if they ignored duplicates, it would be more likely that another duplicate would be submitted ($I=2$); therefore, the best action to take is to ``acknowledge'' the report and mark it as ``Informative.'' In one instance, an interviewee mentioned how duplicate reports came in waves. They described the experience as ``a spree where it's like certainly four or five people at the same time report the same bug again,'' leaving a trail of ``useless reports'' for OSS maintainers to review (P7). Interviewees also indicate receiving duplicates of the same issue even after addressing a prior report ($I=4$).

One interviewee also experienced receiving duplicate reports for a legitimate vulnerability, but the actual reports ranged in quality (i.e., some were more detailed than others). They described a scenario that reflects this, ``the second user who reported the same problem wrote a very nice issue like with code and screenshots, and whatever tests; but I have to mark valid for the first one because that was the first one... I understood that it was a problem, and marked the second one as a duplicate'' (P4). In such cases, the first valid report reviewed would be awarded a bounty, which one interviewee felt was unfair to the bug hunter with the ``better'' report. %\steven{Another usage of "OSS project maintainer" here} 
Another interviewee describes that ``reputation is important because it is the kind of precursor for how trustworthy or capable someone is in the platform, and also it can hurt them'' (P3).

\subsubsection{Feelings when receiving reports}\label{sec:feelings}

As mentioned previously, \textit{project visibility} was the third-most listed benefit and ranked second in significance ($L=13$, $\pi=0.142$), respectively. Interviewees ($I=9$) argue that more criticism comes with more eyes on a project.

Several interviewees brought up the notion of ``software ego'' when receiving bug bounty reports to review ($I=7$). This term can be described in OSS as an OSS maintainer's sense of pride, ownership, or attachment to their project. One interviewee argued having a software ego is ``definitely a very unhelpful mindset because the fact that you put your code out there for people to use, that commits some responsibility... I'm not gonna say you're liable if there's a bug in there, but like at worst, what they report does need to be fixed''  (P9). Other interviewees who brought up ``software ego'' agreed, but one of them still ``felt attacked'' when a bug bounty report was submitted for their project and had to ``not let the ego part of it come into it'' ($I=3$). 
%OSS project maintainers must recognize the positive nature of bug bounty and that ethical hackers are simply trying to discover vulnerabilities. \josh{I'm not sure why this previous sentence is stated so strongly (e.g., ``must recognize the positive nature" and ``simply trying to discover vulnerabilities"). You've already provided substantial evidence that some bug bounty hunters are just trying to cheat their way by maximizing severity score properties and showing evidence of just trying to get money and not improve security.}
%\jessy{feedback addressed - removed}

One interviewee also described how they used to ``get excited'' or ``put everything down'' when they first started receiving bug bounty reports. Over time, that mindset shifted to ``acknowledge then think'' (P12). Five interviewees also mentioned how OSS project maintainers should not ``panic'' and take their time assessing the bug to realize its impact, then creating a plan of resolution. Another interviewee recommended maintainers be upfront with their intentions and let bug hunters know how long it may take to make judgments about the report to lower expectations of a quick review.

\subsection{Vulnerability reproduction, patching, and management}

During bug bounty review, OSS project maintainers argue that being able to reproduce the reported security bug and patching it is critical in ensuring the safety of OSS users. 
These two stages typically occur before public report disclosure, so any obstacles preventing timely reproducibility or deployment fixes will delay notifying affected users.
Alomar et al. \cite{alomar2020bugs} find that a lack of detailed information in reports for reproducing vulnerabilities hinders informed decision making on how to fix them. Further, Akgul et al. \cite{akgul2023BBHperspectives} report that bug bounty hunters realize that their PoCs might not be reviewed or fully understood, leading to disagreements in severity, resulting in lower bounty payouts.

\subsubsection{Reproducibility}\label{sec:reproducibility}

Although \textit{reproduction difficulties} is one of the least listed challenges and ranked eighth ($L=10$, $\pi=0.079$) respectively, some interviewees ($I=5$) argued its importance in aiding effective and timely patching for \textit{efficient development} ($L=7$, $\pi=0.069$). 
Further, although \textit{PoC requirement} is listed the least amount of times in wanted features and is ranked fourth ($L=2$, $\pi=0.106$), there is demand to have such a requirement.

Interviewees argued that the best-detailed reports included a usable PoC ($I=7$), leading to feasible bug reproduction. One argued how having ``the proof of concept makes it a lot easier to actually reproduce the bug... then I don't have to do the paperwork at the end so it's all about [patching]'' (P5), which we present in Section \ref{sec:testing}. With a usable PoC, interviewees described how it helps promote a productive workflow and creates additional time to focus on other development aspects ($I=4$), e.g., features. Further, being able to reproduce the vulnerability easily allows OSS project maintainers to understand potential fixes, which we also provide insight for in Section \ref{sec:testing}.

On the other hand, interviewees expressed concern with reports that did not have an adequate or existing PoC ($I=5$). Some argue that reports should be required to include a working proof-of-concept ($I=3$), which should also have some way of being verified, e.g., by leveraging large-language models ($I=3$). One interviewee described how bug hunters sometimes ``don't really give clear reproduction steps or they give like a corrupt screenshot so you can't even see [the bug]... making it harder to actually chase down where the actual issue is'' (P9). This is especially important because sometimes it turns out to be a false positive ($I=4$). However, as presented in Section \ref{sec:time}, there is limited time to spend on conducting bug bounty report review, let alone on project management ($I=7$).

%\steven{Usage of "OSS project maintainer" here - should it be consistent to be interviewee?}
One interviewee explained how some bug bounty reports are impossible to reproduce because of the availability of hunter-specific tooling. Such tooling is described as pay-to-use, written in another language, or only available in specific geographic regions. Others brought up how it could be helpful for bug bounty platforms to have sandbox environments so that reproducibility is possible in-platform ($I=2$). 

\subsubsection{Fixing and testing}\label{sec:testing}

Although \textit{finding a fix} ($L=7$, $\pi=0.079$) and \textit{testing difficulties} ($L=5$, $\pi=0.113$) are not the most significant challenges respectively, interviewees emphasized the challenges in finding a correct fix for the reported vulnerability ($I=7$). Participants also listed \textit{CI/CD mapping} as a helpful feature ($L=7$, $\pi=0.054$) and automated \textit{third-party integrations} ($L=8$, $\pi=0.074$) as a wanted feature. Together, these features aid in applying fixes and testing during the bug bounty report review process.

Some interviewees described the challenge of determining where in the source code the vulnerability occurs because the bug hunter only provided a PoC and reproduction steps ($I=4$). Another issue interviewees argued is a bug hunter referencing a random line of source code instead of a general area or file where they think the vulnerability may be occurring ($I=2$). One interviewee described this problem with the following, ``you know, it's not hard if you know how this stuff works to search the code base to find a piece of text or a variable name or something and be like, oh, I think this is where the problem is'' (P5). This shows that OSS project maintainers benefit from bug hunter guidance since they are more security-oriented and could provide additional insight, even if it is not a bug fix.

Others discussed knowledge gaps with implementing fixes to reported bugs ($I=2$). In one particular instance, an interviewee described a time when they had so much difficulty trying to implement a fix that they ``limited the functionality of the product'' (P4) as a patch to the reported vulnerability, i.e., by removing a feature. Some interviewees mention how communication gaps, e.g., the bug hunter remaining silent or non-responsive, hinder the ability to create a patch ($I=7$) and that it would be ``nice to work with [bug hunters] during the [patching] process.'' 

Finally, some interviewees argued testing is a significant challenge ($I=2$). In one case, the interviewee described how they received fuzzing-oriented bug bounty reports; however, when it came to testing, ``for [redacted], this usually happens by feeding junk as input and seeing how [redacted] reacts, so the first problem is getting the gist from the junk and making reproducible test cases out of it'' (P14). In other words, even though OSS project maintainers can apply fixes based on bug bounty reports, it can be challenging to create regression test cases for CI/CD purposes ($I=2$).

\subsubsection{Report management}

\textit{Easy report management} was the second most listed helpful feature and ranked fifth ($L=14$, $\pi=0.109$). \textit{More management features} ($L=4$, $\pi=0.104$) and \textit{environmental features} ($L=5$, $\pi=0.106$) were relatively listed less often and are mid-ranked when considering wanted features.

Interviewees argued that having a user-friendly interface and varied features to manage bug reports helped streamline the report-and-resolve process ($I=5$). For instance, one interviewee explained how ``being able to link to our security policy, in some ways, the people are looking at that before they're posting the bug, rather than us having to go back to it afterward, or in some way communicating the expectations for the release cycle'' (P7). 
%\josh{The previous sentence is hard to parse; please include more commas or dashes.} \steven{addressed} 
In this scenario, there is more focus on receiving legitimate reports and clear expectations because of the ability to link a project's security policy. 

Further, interviewees also argued the need for additional report management features ($I=5$), and some mentioned the need to organize duplicate reports and control sorting mechanisms ($I=2$). Others explained how the ability to create and store templates for responding to bug hunters could be handy ``so I don't have to copy-paste it'' ($I=3$). %\josh{The use of ``($I= TBD$)'' is starting to look arbitrary rather than systematic here. First, you mention 5 interviewees without using ``($I= TBD$)'', then 2 participants without including ``($I= TBD$)'' (are they survey participants or interviewees?), and then ``($I= TBD$)'' shows up in the end.} \jessy{will address after @steven is done with updates}

\subsection{Security knowledge}

Despite the importance of actively practicing secure coding practices and being knowledgeable about general security concerns, most developers do not have a security background. A majority of our study participants, as seen in~\autoref{tab:participants}, also lack a security background. 
This section reports a series of challenges while conducting bug bounty report reviews and opportunities where OSS project maintainers were able to learn about aspects of security bugs, e.g., CVSS scoring, and how to go about repairing such bugs through hands-on experience. %\josh{They learned an opportunity? Is that really what you mean?}
Akgul et al. \cite{akgul2023BBHperspectives} find that bug bounty hunters recognize knowledge gap challenges that could result in report misunderstanding.

\subsubsection{Opportunities to learn}

\textit{Learning opportunities} is the second most listed benefit and is ranked fifth ($L=18$, $\pi=0.100$) in importance. OSS project maintainers see a benefit in learning more about (1) ways they can keep their project secure and (2) OSS attack vectors. 

Interviewees argued the significance of learning opportunities that happen as a result of bug bounty report review ($I=5$). 
One interviewee stated, ``I think by now I know most attack vectors, so I know a lot of them, and I know I understand with what security researchers come to me with, and I can I have a clear understanding about it'' (P16). %\josh{Please be mindful of punctuation when quoting. I added more commas to make the previous sentence more readable.} 
This reflects the ability of an OSS project maintainer to develop security knowledge by being involved with the bug bounty report review process. 

Some interviewees were intrigued by the technical measures bug hunters have taken to find vulnerabilities in their project ($I=2$). Further, interviewees have taken the initiative to adopt secure coding practices due to reviewing bug bounty reports ($I=5$). One interviewee, in particular, articulated how they take additional time to review if a feature is necessary or may introduce additional attack surface area. Another interviewee explained how they now apply similar principles to code review when a contributor submits a pull request on their project. These proactive approaches to programming show a sense of adoption post-bug bounty review in a way that increases OSS project security posture.

Other interviewees brought up using Open Worldwide Application Security Project (OWASP) \cite{OWASP} tutorials and documentation as primary references to learn about the most common vulnerabilities ($I=4$), e.g., the OWASP top ten web application security risks \cite{OWASPtop10WAS}. This helped instill confidence when conducting bug bounty review, and interviewees felt they could easily apply gathered knowledge, e.g., ways to prevent SQL injection, to their projects ($I=3$). Further, two interviewees suggest other OSS project maintainers experiment with OSS security tooling against intentionally vulnerable software, e.g., Damn Vulnerable Web Application (DVWA) \cite{DVWA}, to learn more about and become familiar with attacks bug hunters commonly look for.

Interviewees were also at least somewhat familiar with how the CVE process works due to reviewing bug bounty reports ($I=5$), while the majority are still unfamiliar ($I=12$). We present results regarding this topic in Section \ref{sec:cvep}. This is one key instance of how OSS project maintainers have trouble understanding a fundamental process in security awareness, even after having triaged vulnerabilities in their projects with a CVE.

\subsubsection{Knowledge gaps}\label{sec:kgaps}

Security \textit{knowledge gaps} are the third-most listed challenge and ranked ninth ($L=10$, $\pi=0.054$) regarding respective significance. 
Further, the availability of a \textit{security guide and FAQs} is the third most listed and wanted feature but ranks sixth in the Likert-scale study ($L=5$, $\pi=0.101$).
This reflects a need for additional security-related support for newcomers and those without a security background.

Interviewees argued the importance of security knowledge gaps hindering bug assessment and resolution ($I=4$). One stated that ``we just don't even have the knowledge nor the expertise nor the time to look into it and all of a sudden we had people, [bug hunters] were reporting security issues, looking into the program, testing it and dedicating their time'' (P6). %\josh{Again, please be mindful of commas and punctuation when quoting}
%1st I: Networking with security experts
Some interviewees discuss their experience interacting with security experts and how they were able to ``learn a lot [from] just the process alone, even if I didn't have a security background'' ($I=7$). %\josh{Again, please be mindful of commas and punctuation when quoting}
One interviewee believed that being able to shadow someone who reviews bug bounty reports would help them fill in the knowledge necessary to perform effective reviews.

%1st I: Security guide and FAQs
Others discussed the need for starter guides, bug bounty understanding, and security-related FAQs ($I=5$). 
%\josh{Whoa. Now, there are missing commas when not quoting in the previous sentence, which I fixed. :)}\jessy{Thank you Josh :D}
One interviewee specifically states that they ``didn't get into becoming a maintainer because I was security focused'' (P15). In particular, there is concern about setting standards for appropriate behavior from researchers when testing software and interacting with maintainers, understanding the calculation of security metrics such as CVSS scores, and information about CVE creation and qualifications. 

In one instance, an interviewee described live attacks occurring on their system. They were aware that it was a security researcher, so they assumed it was acceptable; however, unbeknownst to the OSS project maintainer, this is considered unethical. This gap reflects how some OSS project maintainers are unaware of boundaries that should not be crossed during bug hunting.

\subsubsection{CVE awareness}\label{sec:cvep}

%1st I: Inverse of familiar with the CVE process
\textit{CVE creation support} is the fourth most listed helpful feature but is ranked second to last ($L=8$, $\pi=0.069$). Though almost the least listed, \textit{security metric calculations} is highly ranked ($L=5$, $\pi=0.133$) as a helpful feature, which directly affects CVE severity. These factors are significant since a majority of interviewees are not familiar with the CVE process ($I=11$). For further context, we recall from~\autoref{tab:participants} that 14 of 17 interviewees have reviewed bug bounty reports resulting in a CVE.

Regarding what qualifies for a CVE, the OSS project maintainers we interviewed have varied perspectives. More than half of interviewees believe reports that result in CVEs have a detailed PoC, a higher severity, and an actual security impact on the software before patching ($I=9$). On the other hand, a majority of interviewees described having little to no understanding of the CVE publishing process ($I=11$), 
%\josh{Do you just mean about the CVE publishing process? Saying ``or not'' sounds awkward.}
despite having reviewed reports that have resulted in CVEs. OSS project maintainers are unclear about what it takes for a reported vulnerability to be assigned a CVE or if a CVE request should be made. In OSS, CVEs are important to upholding the integrity of a project and are used to inform dependent client projects of instances via alerts, e.g., GitHub Dependabot~\cite{dpndbt}, where such software needs to be upgraded because of pre-existing vulnerabilities.
%\josh{What's the benefit of an OSS project maintainer getting a CVE for their project, anyway, especially if the CVE comes out before they can fix it?}

Some interviewees are under the impression that only bug hunters determine whether a report should receive a CVE ($I=2$). One interviewee described a scenario where a bug hunter ``reported what I would consider a security bug with no CVE, I have no idea but that's just the way they disclosed it, so I was like okay cool'' (P13). Another interviewee explained how they did not realize they had a role, or could ``have a say'' (P17), during CVE assignment. This includes actions such as severity adjustment or updating the description by reaching out to MITRE \cite{MITRE}. Some OSS project maintainers are unaware of such actions. 

One interviewee explained, ``I've had a very miserable experience trying to get MITRE to correct incorrect information on existing CVEs, so like I just kind of avoid dealing with all that'' (P10). Although some indicate that they are familiar with the CVE process ($I=6$), others perceive reaching out to CVE authorities as frustrating and actively avoid doing so by making changes to bug reports, i.e., to avoid mistakes before they are routed for CVE review.

\subsection{Working alongside bug hunters}

OSS project maintainers identify that communication and collaboration are crucial to the success of report review. Alomar et al. \cite{alomar2020bugs} found that bug bounty hunters can build productive relationships with security teams, allowing them to collaborate on fixing vulnerabilities and receiving prompt feedback. Although prior work suggests that lack of communication is one of the more challenging components for bug hunters \cite{maulani2023case-study, akgul2023BBHperspectives, li2022relational, ellis2022bounty}, OSS project maintainers felt as if it was one of the least challenging aspects of reviewing bug bounty reports---confirming opposing views of bug bounty hunters and reviewers.

\subsubsection{Communication}

\textit{Networking opportunities} was the fourth least-listed benefit and the lowest ranked ($L=12$, $\pi=0.061$). Related to networking, having a \textit{communication system} was also listed less often but ranked fourth in helpfulness ($L=8$, $\pi=0.106$). Surprisingly, \textit{lack of communication} was listed as the second least often and the least challenging ($L=6$, $\pi=0.038$). 

Several interviewees argued for security expert importance during the review process ($I=7$), two of which have even maintained communication with bug hunters post-disclosure for future assistance with vulnerability finding and repair. We present results on this topic in Section \ref{sec:colab}. One interviewee argued that communicating with security experts was important because ``they explained to us not just how to reproduce but also the impact'' (P12), challenges reported in Section \ref{sec:reproducibility}. Another interviewee explained how having access to security experts is needed in OSS bug bounty review because ``as maintainers we kind of started at the very beginning, we have so many different hats on and have come into the maintainer-ship through many different avenues'' (P11), referencing knowledge gaps reported in Section \ref{sec:kgaps}.

Other interviewees emphasized the importance of having an in-platform communication system for threaded discussions with bug hunters ($I=5$). Further, they believe it is important for tracked conversations, real-time communication, and having the ability to provide comments. One interviewee explained how they would prefer to keep communication isolated to email. In contrast, some interviewees explicitly stated that they would rather keep bug bounty report exchanges limited to a central platform ($I=4$). Others have mixed views of what method of communication is best for vulnerability resolution but would prefer to have them somewhere other than in the public eye ($I=3$). One interviewee explained how they were fine with a bug hunter opening a GitHub issue for a vulnerability because their project is relatively small, but understands why more popular projects would not want this. 
%\josh{As in ``more popular projects'', unless that small project really is unpopular.}

Few interviewees argued that lack of communication is a challenge during bug bounty review ($I=3$). One interviewee described how lack of communication is a problem when they cannot get hunter input for locating the vulnerability in source code or during patch development, both of which details are reported in Section \ref{sec:testing}. Another interviewee described unfortunate experiences, initially reported in Section \ref{sec:exp}, where bug hunters disclosed vulnerabilities before a patch was released. ``I've also had times that somebody released one without even talking to me'' (P9). While our studies' results do not emphasize there is a lack of communication, the consequences can be severe and compromise trust in bug bounty programs for OSS.

\subsubsection{Collaboration}\label{sec:colab}

Recall, \textit{networking opportunities} was the fourth least-listed benefit and the lowest ranked ($L=12$, $\pi=0.061$). In relation, OSS project maintainers express a need for \textit{collaboration features} ($L=2$, $\pi=0.163$). Although the least listed compared to other wanted features, to request reviews from security experts or other maintainers, it is ranked as the most wanted feature.

Interviewees see bug bounty programs as an avenue to expand their collaboration network ($I=6$). In some cases, the bug hunter becomes an active contributor to the project for which they originally submitted a bug bounty report ($I=2$). ``We know that they're genuinely interested in the improvement of the project, and we really appreciate that kind of contribution'' (P17). The same interviewees view bug hunters as contributors, regardless of their decision to continue contributing, but emphasized the importance of their security background and role in helping improve OSS project security posture. 

Some prefer bug hunters to work directly alongside OSS project maintainers to review the patch before deployment and disclosure ($I=5$); therefore, they request bug bounty platforms have a feature that can easily facilitate such collaboration, e.g., adding other maintainers to the discussion. Further, one interviewee argues how it would be beneficial to have a feature that facilitates peer reviews, allowing for collaborative assessment and verification of reported vulnerabilities, and believes this could enhance the overall quality and accuracy of the bug bounty evaluation process.

%% file: content/original_table.tex
 \begin{table*}[th!]
 \centering
 \begin{adjustbox}{width=\textwidth,totalheight=4.95in}
 \begin{tabular}{crlrrl}
 \toprule
 %\midrule
 & \multicolumn{1}{r}{\textbf{Characteristic}} & \multicolumn{1}{l}{\textbf{Description}} & \multicolumn{1}{r}{\textbf{RW}} & \multicolumn{1}{r}{\textbf{L}} & \multicolumn{1}{l}{$\bm{\pi}$}\\
 \midrule
 \parbox[t]{3mm}{\multirow{10}{*}{\rotatebox[origin=c]{90}{\textbf{Benefits of BBR Review}}}}
   & \textbf{Private disclosure} & Confidential communication of vulnerabilities to project maintainers. & \cite{beretas2023analysis, subramanian2020marketplaces, kuehn2014insttitutional} & 4 & 0.155 \\
   & \cellcolor{grey1} \textbf{Project visibility} & \cellcolor{grey1}``Many eyes'' are available to help review a project. & \cellcolor{grey1}\cite{maillart2017eyes, alexopoulos2021vuln-reporters} & \cellcolor{grey1}13 & \cellcolor{grey1}0.142 \\
   & \textbf{Detailed reports} & Receiving thorough bug reports to help timely patching. & & 8 & 0.138 \\
   & \cellcolor{grey1}\textbf{Improved security posture} & \cellcolor{grey1}Overall project defense, minimizing vulnerability risks. & \cellcolor{grey1}\cite{zhao2015web-vulns, walshe2022cvd, atefi2023chromium, maulani2023case-study, bhushan2022dynamics, hendrick2023openssf, wermke2022qual-OSSP, zhao2016crowdsourced} & \cellcolor{grey1}33 & \cellcolor{grey1}0.124 \\
   & \textbf{Learning opportunities} & The ability to learn about threats and mitigations. & \cite{akgul2023BBHperspectives, ellis2022bounty} & 18 & 0.100 \\
   & \cellcolor{grey1}\textbf{Benefits OSS users \& trust} & \cellcolor{grey1}Transparent handling of security issues builds trust with users. & \cellcolor{grey1}\cite{wermke2022qual-OSSP} & \cellcolor{grey1}13 & \cellcolor{grey1}0.077 \\
   & \textbf{Delegated bounties} & Bounty payments come from and are handled by an external entity. & \cite{hou2023temporal, finifter2013vuln-rewards, akgul2023BBHperspectives} & 9 & 0.070 \\
   & \cellcolor{grey1}\textbf{Efficient development} & \cellcolor{grey1}Additional time to focus on other development aspects. & \cellcolor{grey1} & \cellcolor{grey1}7 & \cellcolor{grey1}0.069 \\
   & \textbf{Outsourced work} & Outsourced red-teaming of OSS project and community involvement. & \cite{walshe2020empiricalBBPs} & 4 & 0.064 \\
   & \cellcolor{grey1} \textbf{Networking opportunities} & \cellcolor{grey1}The ability to obtain security expert input and build connections. & \cellcolor{grey1}\cite{alexopoulos2021vuln-reporters, huang2016multiple, ellis2022bounty} & \cellcolor{grey1}\cellcolor{grey1}12 & \cellcolor{grey1}0.061 \\
  \midrule
 \parbox[t]{3mm}{\multirow{12}{*}{\rotatebox[origin=c]{90}{\textbf{Challenges in BBR Review}}}}
   & \textbf{Diverted focus} & Hunters are focused on money or CVEs, not software security. & \cite{hou2023temporal, finifter2013vuln-rewards, laszka2016banish, maillart2017eyes, ellis2022bounty} & 12 & 0.141 \\
   & \cellcolor{grey1}\textbf{Pressure to review} & \cellcolor{grey1}Reporters pushing to get their findings approved fast. & \cellcolor{grey1} & \cellcolor{grey1}12 & \cellcolor{grey1}0.113 \\
   & \textbf{Testing difficulties} & Difficulty creating test cases for applied fixes. & & 5 & 0.112 \\
   & \cellcolor{grey1}\textbf{Overstated severity} & \cellcolor{grey1}Dealing with reports that do not reflect the bug's true severity. & \cellcolor{grey1} & \cellcolor{grey1}9 & \cellcolor{grey1}0.096 \\
   & \textbf{Low-quality or spam} & Dealing with low-quality or spam reports. & \cite{walshe2022cvd, sridhar2021hackerone, shafigh2021invalid, laszka2016banish, zhao2020control, zhao2016crowdsourced} & 22 & 0.094 \\
   & \cellcolor{grey1}\textbf{Validating impact} & \cellcolor{grey1}Evaluating the practical security impact of a report. & \cellcolor{grey1} & \cellcolor{grey1}11 & \cellcolor{grey1}0.085 \\
   & \textbf{Finding a fix} & Issues with finding and implementing fixes. & & 7 & 0.079 \\
   & \cellcolor{grey1}\textbf{Reproduction difficulties} & \cellcolor{grey1}Challenging to reproduce reported bugs for effective understanding. & \cellcolor{grey1} & \cellcolor{grey1}10 & \cellcolor{grey1}0.079 \\
   & \textbf{Time-consuming} & Reviewing bug bounty reports takes up a lot of time. & \cite{walshe2022cvd} & 10 & 0.066 \\
   & \cellcolor{grey1}\textbf{Knowledge gaps} & \cellcolor{grey1}Gaps in security knowledge hinder bug assessment and resolution. & \cellcolor{grey1}\cite{hendrick2023openssf} & \cellcolor{grey1}10 & \cellcolor{grey1}0.054 \\
   & \textbf{Duplicate reports} & Dealing with multiple identical or similar bug reports. & \cite{nagwani2021slr-dupes, ellis2022bounty, zhao2016crowdsourced} & 6 & 0.043 \\
   & \cellcolor{grey1}\textbf{Lack of communication} & \cellcolor{grey1}Inadequate communication between reporters and maintainers. & \cellcolor{grey1}\cite{maulani2023case-study, akgul2023BBHperspectives, li2022relational, ellis2022bounty} & \cellcolor{grey1}6 & \cellcolor{grey1}0.038 \\
 \midrule
   \parbox[t]{3mm}{\multirow{9}{*}{\rotatebox[origin=c]{90}{\textbf{Helpful Features}}}}
   & \textbf{Incentivized reporting} & Bug hunters receive points for reporting valid vulnerabilities. & & 4 & 0.167 \\
   & \cellcolor{grey1}\textbf{Security metric calculations} & \cellcolor{grey1}Assistance in calculating security metrics (e.g., CVSS score). & \cellcolor{grey1} & \cellcolor{grey1}5 & \cellcolor{grey1}0.133 \\
   & \textbf{Feedback capabilities} & The ability to provide feedback and make report adjustments. & & 6 & 0.122 \\
   & \cellcolor{grey1}\textbf{Reducing project costs} & \cellcolor{grey1}The existence of financial incentives from an external entity. & \cellcolor{grey1}\cite{walshe2020empiricalBBPs} & \cellcolor{grey1}12 & \cellcolor{grey1}0.121 \\
   & \textbf{Secure disclosure} & Receiving private bug reports and publishing them after patching. & \cite{beretas2023analysis, badash2021blockchain} & 15 & 0.118 \\
   & \cellcolor{grey1}\textbf{Easy report management} & \cellcolor{grey1}User-friendly interface and varied features to manage bug reports. & \cellcolor{grey1} & \cellcolor{grey1}14 & \cellcolor{grey1}0.109 \\
   & \textbf{Communication system} & Comments and messaging are available during the review process. & \cite{maulani2023case-study} & 8 & 0.106 \\
   & \cellcolor{grey1}\textbf{CVE creation support} & \cellcolor{grey1}Streamlined and assisted process of creating CVEs. & \cellcolor{grey1} & \cellcolor{grey1}8 & \cellcolor{grey1}0.069 \\
   & \textbf{CI/CD mapping} & Integrating basic CI/CD details into bug reports. & & 7 & 0.054 \\
  \midrule
   \parbox[t]{3mm}{\multirow{9}{*}{\rotatebox[origin=c]{90}{\textbf{Wanted Features}}}}
   & \cellcolor{grey1}\textbf{Collaboration features} & \cellcolor{grey1}The ability to request reviews from security experts or other maintainers. & \cellcolor{grey1}\cite{maulani2023case-study} & \cellcolor{grey1}2 & \cellcolor{grey1}0.163 \\ 
   & \textbf{Reputation system} & Additional ways to measure and assign reputation to bug hunters. & \cite{ohare2020game} & 4 & 0.145 \\
   & \cellcolor{grey1}\textbf{More financial incentives} & \cellcolor{grey1}E.g., rewards for bug fixes, bonuses for high-quality reports. & \cellcolor{grey1}\cite{hou2023temporal, finifter2013vuln-rewards, gersbach2023decentralized, ellis2022bounty} & \cellcolor{grey1}4 & \cellcolor{grey1}0.145 \\
   & \textbf{PoC requirement} & Reports will be required to include a working proof-of-concept (PoC). & & 10 & 0.106* \\
   & \cellcolor{grey1}\textbf{Environmental features} & \cellcolor{grey1}E.g., support for in-built code review, rich text comments. & \cellcolor{grey1} & \cellcolor{grey1}5 & \cellcolor{grey1}0.106* \\
   & \textbf{More management features} & E.g., report sorting, the ability to link duplicate reports. & & 4 & 0.104 \\
  & \cellcolor{grey1}\textbf{Security guide \& FAQs} & \cellcolor{grey1}Guides for bug bounty understanding and security-related FAQs. & \cellcolor{grey1}\cite{hendrick2023openssf} & \cellcolor{grey1}5 & \cellcolor{grey1}0.101 \\
  & \textbf{Third-party integrations} & \underline{Automated} integration of CI/CD features for additional awareness. & & 8 & 0.074 \\
  & \cellcolor{grey1}\textbf{Scheduled disclosure} & \cellcolor{grey1}The ability to schedule when the report will be disclosed. & \cellcolor{grey1} & \cellcolor{grey1}2 & \cellcolor{grey1}0.056 \\
  \bottomrule
 \end{tabular}
 \end{adjustbox}
 \caption{40 characteristics identified from the listing survey study divided into four categories as a result of bug bounty report (\textbf{BBR}) review. Acronyms listed represent, respectively: existing related work (\textbf{RW}), the number of listing (\textbf{L}) survey study mentions, and worth estimates ($\bm{\pi}$), for ranking responses, from the Likert-scale survey study. * indicates a tie.}
    \label{tab:bigasstable}
 %\end{centering}
 \end{table*}

%% file: content/discussion.tex
\section{Discussion}\label{sec:discussion}

In this section, we list recommendations for OSS-focused bug bounty programs, platforms, and OSS project maintainers to alleviate prominent challenges we discovered and support and expand the benefits participants expressed as the most helpful. Further, we present potential research directions to support OSS project maintainers and bug hunters during the bug bounty lifecycle.

\subsection{Challenges and their implications}

OSS project maintainers mentioned low-quality or spam reports the most in our listing study and rated a diverted focus on money or CVEs as the most challenging when conducting bug bounty report review, both of which were discussed by a majority of the interviewees. Other relevant aspects include overstated severity and duplicate reports. 

Low-quality or spam reports can overwhelm OSS maintainers and take away from completing project tasks or resolving valid vulnerability reports. Bug bounty platforms should implement mechanisms that can automatically identify low-quality, spam, and duplicate reports. Bug bounty programs should strive to implement a reward structure that considers monetary incentives and recognizes and rewards contributors for their impact, which is especially important in OSS. Bug bounty platforms should provide OSS project maintainers the ability to rate hunters with reputation scores. %\josh{This paragraph may be a good place to mention how OSS project maintainers can rate hunters in terms of reputation scores.}

Other prominent challenges OSS project maintainers faced during bug bounty review are pressure, knowledge gaps, and bug reproduction difficulties. Collectively, these aspects may lead to overlooking critical vulnerabilities or applying inadequate patches because of being overwhelmed and having an insufficient understanding of bugs. Bug bounty platforms should have a guide for bug hunter best practices, e.g., detailed steps for reproducing vulnerabilities and usable support mechanisms so that reviewers can seek assistance when faced with challenges regarding pushiness or knowledge gaps. Bug bounty programs should strive to encourage bug hunters to understand the workload and background of OSS project maintainers and to be patient during the resolution process, e.g., by having staff oversee or mediate interactions. Further, bug bounty platforms should incentivize hunters to maintain positive reputation scores so OSS project maintainers can scope reliability, which is further supported by our finding about how OSS project maintainers rate diverted focus, i.e., hunters are focused on bounties and CVEs, as the most challenging to deal with when conducting bug bounty report review. %\josh{Maybe this is a good paragraph to mention a reputation score for hunters given by OSS maintainers again.}

\subsection{Benefits and their implications}

Private disclosure, improved security posture, and project visibility were amongst the most important benefits for maintainers after conducting bug bounty report review. 

OSS project maintainers should communicate the positive impact of bug bounty programs on OSS so other projects reluctant to participate can be intrigued by the benefits of receiving a report from a third-party program. 
Further, we discourage OSS project maintainers from guiding users to submit vulnerabilities using public GitHub issues---which is, in fact, an already discouraged practice \cite{gh_issues}---and should instead take advantage of private vulnerability reporting \cite{gh_pvr,gh2023private} regardless of project size. 
Bug bounty platforms should implement incentive structures that reward bug hunters not only for vulnerability discovery but also for detailed or easy-to-reproduce bug reports and active collaboration. Bug bounty programs should consider implementing features that allow reviewers to visualize security impact, e.g., a chart of patched bugs alongside severities, to associate report resolution with positivity over time.

Bug bounty report review also presented various beneficial learning opportunities for OSS project maintainers. OSS project maintainers should proactively participate in the review process to learn about common vulnerabilities and effective mitigation strategies, and they should also do so with hands-on experience by constructing patches for their projects. Bug bounty programs should consider collaborating with security experts to provide training sessions tailored for OSS project maintainers and facilitate collaboration opportunities between maintainers and security experts. Bug bounty platforms should consider providing a sandbox environment whenever possible so reviewers can simulate reported bugs, allowing immediate vulnerability reproduction and collaboration functionality so that fixes can be easily co-developed by bug hunters and report reviewers. Further, bug bounty platforms should ensure maintainers are being properly incentivized for their efforts, especially since such maintainers typically lack adequate time and resources.

\subsection{Future work}

Most interviewees expressed generally positive and hopeful attitudes toward integrating large language models (LLMs) during the bug bounty lifecycle ($I=13$). Some even brought up past experiences from the perspective of a bug hunter. Further, there are security education opportunities to explore that could help the review process and understanding of vulnerability disclosure, e.g., redesigning the CVE creation support process to be more informative so that bug bounty report reviewers can find it more helpful.

We encourage researchers to investigate further how LLMs can be leveraged to (1) help bug hunters create more effective reports, (2) help reviewers speed up reviews without losing quality and context, and (3) help reviewers generate patches for vulnerabilities. 
One potentially promising direction researchers can explore is harnessing LLMs with content from OSS bug bounty reports to enhance current state-of-the-art LLM techniques used for OSS vulnerability discovery, PoC exploit and test generation, severity determination, and repair.
We also encourage researchers to develop practical educational resources accessible to developers without a security background that support the review process, further investigate other knowledge gaps this subset of developers have about the vulnerability disclosure process, and develop methods to help close such gaps.

One multi-dimensional challenge our studies raise is ensuring that fixes proposed by a hunter do not introduce regressions, which is difficult because the CI/CD systems on GitHub are not accessible to or reusable by hunters and may be resolved through automation or infrastructure support so that PoCs can be included as tests into the CI/CD systems of projects.
Our Likert study highlights that pressure to review combined with the difficulty of creating test cases for applied fixes, overstated severity, validating a report’s impact, and finding a fix together emphasize the need to automate the severity determination, fixing of vulnerabilities, and verifying the fixes to ease the burden of maintainers. As column 2 of \ref{tab:bigasstable} shows, we found little related work in this space. Automated program repair and efficient regression testing remain open research challenges.

%\subsection{Recommendations}

%\subsection{Security education}

%\subsection{Future work for supporting the bug bounty lifecycle}

%% file: content/conclusion.tex
\section{Conclusion}\label{sec:conclusion}
In this paper, we conduct a mixed-methods study, i.e., two surveys and semi-structured interviews, to investigate OSS vulnerability disclosure practices by systematically identifying and quantifying the factors that affect how OSS project maintainers conduct bug bounty report review. OSS project maintainers find private disclosure and project visibility to be the most beneficial aspects of bug bounty report review, while hunters' diverted focus and pressure to review bug bounty reports to be the most challenging. We also find that the most helpful features are incentivized reporting and security metric calculation assistance, while collaboration features and additional hunter reputation features are the most wanted capabilities on bug bounty platforms.
%Our study includes the following key findings: 
%(1) current review rates cannot keep up with growing unreviewed security advisories; 
%(2) CVEs missing from the NVD prevent alerts from being sent to affected GitHub projects and external advisory databases; 
%(3) detailed PoCs and multiple sources of evidence are crucial for receiving a CVE and should be kept private before disclosure; 
%(4) a majority of bug bounty reports and security advisories analyzed have invalid reasons for missing CVEs, hindering vulnerability awareness; and 
%(5) a majority of projects analyzed do not show past patched vulnerabilities or enable private vulnerability reporting, leaving many gaps in proper disclosure. 
%Based on our findings, we have provided suggestions and actionable items to improve the OSS ecosystem security posture.